\def\updated{Last update by meb/dja 21 March 2003}
\documentclass[twocolumn]{mn2e}

\def\beq{\begin{equation}}
\def\eeq{\end{equation}}
\def\bey{\begin{eqnarray}}
\def\eey{\end{eqnarray}}

\def\TJ{{T_{\rm J}}}
\def\TS{{T_{\rm S}}}
\def\TU{{T_{\rm U}}}
\def\TN{{T_{\rm N}}}
\def\TP{{T_{\rm P}}}
\def\Msun{{M_\odot}}
\def\RH{{R_{\rm H}}}
\def\MP{{M_{\rm P}}}

\def\aP{{a_{\rm P}}}

\def\and{{\rm and\ }}
\def\be{\begin{equation}}
\def\ee{\end{equation}}

\def\spose#1{\hbox to 0pt{#1\hss}}
\def\lta{\mathrel{\spose{\lower 3pt\hbox{$\sim$}}
    \raise 2.0pt\hbox{$<$}}}
\def\gta{\mathrel{\spose{\lower 3pt\hbox{$\sim$}}
    \raise 2.0pt\hbox{$>$}}}

 \input epsf

\title[Populations of Comet-Like Bodies]
      {The populations of comet-like bodies in the Solar system}

\author[J. Horner et al.]
       {J. Horner$^1$, N.W. Evans$^{1,2}$, M.E. Bailey$^3$ \& D.J. Asher$^3$\\
       $^1$ Theoretical Physics, 1 Keble Rd, Oxford, OX1 3NP \\
       $^2$ Institute of Astronomy, Madingley Rd, Cambridge CB3 0HA \\
       $^3$ Armagh Observatory, College Hill, Armagh, BT61 9DG}
\date{\updated} 
\pagerange{\pageref{firstpage}--\pageref{lastpage}}
\pubyear{}

\begin{document}
\maketitle
\label{firstpage}

\begin{abstract} A new classification scheme is introduced for comet-like
bodies in the Solar system. It covers the traditional comets as well as
the Centaurs and Edgeworth-Kuiper belt objects.  At low inclinations,
close encounters with planets often result in near-constant perihelion or
aphelion distances, or in perihelion-aphelion interchanges, so the minor
bodies can be labelled according to the planets predominantly controlling
them at perihelion and aphelion. For example, a JN object has a perihelion
under the control of Jupiter and aphelion under the control of Neptune,
and so on.  This provides 20 dynamically distinct categories of outer
Solar system objects in the Jovian and trans-Jovian regions. The Tisserand
parameter with respect to the planet controlling perihelion is also often
roughly constant under orbital evolution. So, each category can be further
sub-divided according to the Tisserand parameter.

The dynamical evolution of comets, however, is dominated not by the
planets nearest at perihelion or aphelion, but by the more massive
Jupiter. The comets are separated into four categories -- Encke-type,
short-period, intermediate and long-period -- according to aphelion
distance.  The Tisserand parameter categories now roughly correspond
to the well-known Jupiter-family comets, transition-types and
Halley-types. In this way, the nomenclature for the Centaurs and
Edgeworth-Kuiper belt objects is based on, and consistent with, that
for comets.  Given the perihelion and aphelion distances together with
the Tisserand parameter, our classification scheme provides a
description for any comet-like body in the Solar system.  The
usefulness of the scheme is illustrated with examples drawn from
numerical simulations and from the present-day Solar system.
\end{abstract}


\section{Introduction}

There is a need to improve the taxonomy for comet-like bodies in the Solar
system.  First, as a result of recent discoveries, the populations of
comet-like bodies in the Solar system are known to be much more extensive
than previously thought. The last decade has seen the discovery of more
than 600 trans-Neptunian objects, beginning with (15760)\,1992\,QB$_1$
(Jewitt \& Luu 1993), as well as the identification of $\sim$100 Centaurs
(e.g.\ Scotti 1992) to supplement the early serendipitous finding of
(2060)\,Chiron (Kowal 1979).  Moreover, many unusual objects have now been
found circulating essentially within the known planetary system, for
example (5335)\,Damocles (e.g.\ Asher et al.\ 1994), and the `asteroids'
(4015)\,Wilson-Harrington = 107P/Wilson-Harrington, (7968)\,Elst-Pizarro =
133P/Elst-Pizarro and (2060)\,Chiron = 95P/Chiron.  These objects blur the
traditional clear distinction between comets and asteroids, either on
physical or dynamical grounds.

Secondly, the spatial distribution of objects in orbits largely beyond
Neptune, but with aphelion distances $Q$ much less than the conventional
inner edge of the dynamically active Oort cloud ($Q \! \ll \!
20\,000\,$au, say), is believed to comprise a flattened disc or belt-like
structure.  These objects have orbits that in some cases extend hundreds
of astronomical units (au) from the Sun, and although resembling asteroids
in appearance are widely thought to be cometary in composition.  This
trans-Neptunian zone includes the region referred to as the Kuiper or
Edgeworth-Kuiper belt (EKB), and probably merges in the form of an
extended trans-Neptunian disc into the inner core of the Oort cometary
cloud.  Trans-Neptunian objects are a possibly significant source of
low-inclination `Jupiter-family' comets (Quinn, Tremaine \& Duncan 1990),
again muddying the traditional distinction between comets and asteroids.

In fact, the Centaur population appears to represent an important link
between trans-Neptunian objects and Jupiter-family comets (Dones, Levison
\& Duncan 1996, Stern \& Campins 1996, Levison \& Duncan 1997).  Centaur
orbits are typically planet-crossing and have relatively short dynamical
lifetimes ($\sim$$10^6$\,yr). Chiron, which is one of a number of
exceptionally large minor bodies with perihelia close to or within the
orbit of Saturn, exhibits cometary activity (e.g.\ Luu \& Jewitt 1990) and
even has a periodic comet designation: 95P/Chiron. At least one recently
discovered comet, namely C/2001\,T4 (NEAT), has a very similar Chiron-like
Centaur orbit. Taken together, the evidence suggests a picture in which
comets and distant minor planets are dynamically reprocessed from one
population to another, for example from the Edgeworth-Kuiper belt, through
Centaurs to Jupiter-family comets, and so on.  Similarly, it is possible
for objects to be scattered out of the main asteroid belt and evolve into
similar areas. So, it is possible that some Centaurs may be rocky or
asteroidal, others may be icy or cometary. This is analagous to the
examples given by Fern\'andez, Gallardo \& Brunini (2002) of mixing
between the Jupiter-family and near-Earth asteroid populations, and again
highlights both the difficulty of separating physically distinct
populations of objects using purely dynamical characteristics, and the
need for a unified dynamical classification scheme capable of describing
the full range of observed orbital types.

In parallel with these observational advances, the march of computer
power now ensures that the orbits of most Solar system bodies can be
routinely integrated for millions of years. Given the wealth of
simulation data and the diversity of new discoveries, the taxonomy of
Solar system objects assumes great importance. For a classification
scheme to be useful, it should allow us to place objects with similar
physical or dynamical characteristics in sets, and to examine the bulk
statistics of the sets in detail.  However, in making sense of the
complicated evolutionary pathways followed by objects such as
short-period comets and Centaurs, the historical classification scheme
seems obsolete.  A new taxonomy should clarify the principal dynamical
paths followed by different classes of object through interplanetary
space, and facilitate the definition of dynamical lifetimes and the
flux and transfer probabilities of objects from one dynamical class to
another.

Traditionally, comets have been classified as short-period or
long-period according to whether their periods of revolution $P$ are
less than or greater than 200~years.  This unphysical division
(Levison 1996) has frequently been justified in terms of the time over
which comets have been observed with scientific precision (e.g.\
Weissman 2001). But, the canonical upper limit for `periodic' comets
has now been spectacularly broken by the return of 153P/Ikeya-Zhang
with an orbital period in excess of 350~years, and also by recent
discoveries such as asteroid 2002\,RP$_{120}$ with a period of about
420\,yr.

Centaurs are conventionally defined as asteroids or cometary nuclei
circulating largely between the orbits of Jupiter and Neptune and usually
crossing the orbit of at least one giant planet (e.g.\ Jedicke \& Herron
1997, Larsen et al.\ 2001).  Jedicke \& Herron's definition restricts
Centaur semi-major axes to less than that of Neptune (i.e.\ $a\!\lta
30\,$au), and therefore allows objects defined as Centaurs to pass
significantly beyond Neptune's orbit into the Edgeworth-Kuiper belt.  A
straightforward extension of this definition would allow Centaurs to be
regarded simply as a continuation of the `scattered disc objects', perhaps
even encompassing those resonant objects that are not protected in the
long term from close approaches to Neptune.  The scattered disc was the
name introduced by Duncan \& Levison (1997) following the discovery of
1996\,TL$_{66}$ (Luu et al.\ 1997) to describe a population of relatively
high-inclination, high-eccentricity trans-Neptunian objects. The idea that
Centaurs and the scattered disc may be drawn from the same underlying
population has been advocated recently both by Marsden (1999a) and by
Emel'yanenko, Asher \& Bailey (2003).

This paper introduces a new classification scheme for comet-like bodies of
the Solar system.  There is much evidence from previous work (e.g.\
Kazimirchak-Polonskaya 1972; Kres\'ak 1972, 1980, 1983; Everhart 1977;
Rickman \& Froeschl\'e 1988; Manara \& Valsecchi 1991; Dones et al.\ 1996;
Levison 1996) that the evolution of comets and Centaurs often takes place
under the control of one or another major planet, usually with
near-constant aphelion or near-constant perihelion, or via
aphelion-perihelion interchanges.  This motivates us to consider the
planets controlling the perihelion $q$ and aphelion $Q$ as fundamental to
the classification of any comet-like bodies. There is also evidence that
low-inclination objects tend to evolve inwards, whereas the reverse is
true for high-inclination objects. This suggests that we use the
inclination -- or better still, the Tisserand parameter $\TP$ with respect
to the controlling planet -- as a third criterion.   The Tisserand
parameter is an approximate constant of the motion, at least when the
orbital perturbations are dominated by a single planet. Given the set of
values ($q,Q, \TP$), our new taxonomy provides an instantaneous
classification for any comet-like body in the Solar system.

\section{An Outline of the Classification Scheme}

\subsection{The zone of control}

The Hill radius is defined as
\begin{equation}
\RH = a \left[ {\MP \over 3\Msun} \right]^{1/3}, 
\end{equation}
where $\MP$ is the mass of the planet, $\Msun$ is the mass of the Sun
and $a$ is the semi-major axis of the orbit of the planet about the
Sun (e.g.\ Murray \& Dermott 1999). It corresponds to the position of
the Lagrange points in the restricted three-body problem (e.g.\ Arnold
et al.\ 1987) and so marks the largest distance at which the planet
may possess a moon.  A Solar system object can be classified according
to the planets which have the greatest effect at perihelion and at
aphelion.  For typical eccentricities, a perihelion lying 4 or 5 Hill
radii beyond an inner planet is often sufficiently distant for the
dynamics to be controlled by the outer planet.  So, we conclude that
for a planet to control an object, then the perihelion or aphelion
must lie closer than a distance of $\approx$3 Hill radii to the
planet's orbit (see e.g., Charnoz, Th\'ebault \& Brahic 2001).

In our classification scheme, each planet's zone of control extends to
$3k$ Hill radii, with the parameter $k \simeq 1$.  In this way, the
Solar system is banded into concentric zones of control of first
Jupiter, then Saturn, Uranus and Neptune.  Over the next few million
years the mean semi-major axes and maximum aphelia of the four major
planets Jupiter, Saturn, Uranus and Neptune are (5.2, 5.5), (9.6,
10.4), (19.2, 20.8) and (30.1, 31.0)\,au respectively (Applegate et
al.\ 1986). Three times their Hill radii are 1.1, 1.3, 1.4
and 2.3\,au respectively.  These distances provide a rough basis for
estimating the respective boundaries between the zones of control. For
example, for Jupiter to control the object at perihelion, the
perihelion must lie between 4 and 6.6 au approximately. If the
aphelion also lies in the range, then we classify the minor body as a
J object.  For Saturn to control the object at aphelion, then $6.6< Q
<12.0$ when the object is denoted by JS. We will give detailed
examples shortly, but the general idea should now be obvious to the
reader.

\subsection{The Tisserand parameter}

\label{sec:tisserand}

The Tisserand parameter is an approximation to the Jacobi constant,
which is an exact integral of motion in the circular restricted
three-body problem. It is defined as (e.g.\ Murray \& Dermott 1999)
\begin{equation} 
\TP = {\aP \over a} + 2\cos i \sqrt{ {a(1-e^2) \over \aP}}  
\end{equation}  
where $\aP$ is the planet's semi-major axis.  

\begin{figure}
\epsfysize=10cm \centerline{\epsfbox{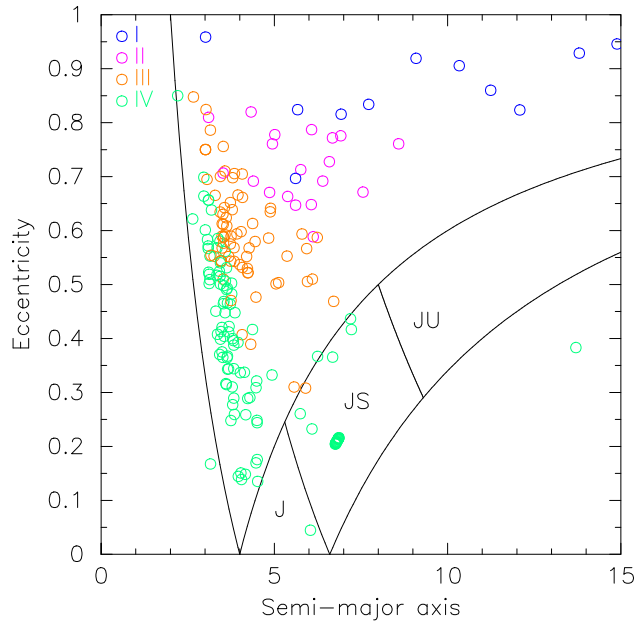}}
\caption{Plot of eccentricity versus semi-major axis for
short-period comets in Marsden \& Williams (1999).  All objects are
colour-coded according to their Tisserand parameter classes with
respect to Jupiter. Most of the comets lie in the SP category,
leftward of the $q = 4.0 $ au line. To the right of that line, the
boundaries of the J, JS and JU categories are marked.}
\label{fig:cometpop}
\end{figure}

Ignoring the positive orbital eccentricity of the planet, objects with
$\TP$ greater than 3 are in principle confined to regions wholly interior
or wholly exterior to the planet's orbit.  There is a long history of
usage of the Tisserand parameter with respect to Jupiter $\TJ$ to divide
the cometary families (e.g., Kres\'ak 1972, 1980, 1982, 1983, 1985; Vaghi
1973 a,b; Carusi \& Valsecchi 1987; Levison \& Duncan 1994; Levison
1996).  Fig.~\ref{fig:cometpop} shows all short-period comets in the
plane of semi-major axis $a$ and eccentricity $e$. The comets have been
colour-coded according to their Tisserand parameter with respect to
Jupiter. Motivated by these plots, we propose a four-fold division
according to Tisserand parameter as indicated in Table~\ref{T1}.

\begin{table}
\begin{center}
\begin{tabular}{cc}\hline
 & \\
Tisserand  & Tisserand \\
Class      & Parameter \\
 & \\
\hline
 & \\
Class I    & $\phantom{2.5 \lta} \TP \lta 2.0$ \\
Class II   & $2.0 \lta \TP \lta 2.5$ \\
Class III  & $2.5 \lta \TP \lta 2.8$ \\
Class IV   & $2.8 \lta \TP \phantom{\lta 2.5} $ \\
 & \\
\hline
\end{tabular}
\caption{Definition of Tisserand parameter classification scheme. The
boundary near $\TP \simeq 2.8$ marks the limit above which it is
impossible for an object to be directly ejected in a single encounter.}
\label{T1}
\end{center}
\end{table}

In our classification scheme, we use the Tisserand parameters with respect
to the major planets ($\TJ, \TS, \TU$ and $\TN$ with obvious notation) to
classify objects whose perihelion lies within the zone of the planet's
control. So, for example, an SU$_{\rm IV}$ object is a minor body whose
perihelion is controlled by Saturn and aphelion by Uranus and whose
Tisserand parameter with respect to Saturn is larger than approximately
2.8.  Here and henceforth, we always denote the Tisserand parameter class
as a subscript.

\section{Comets}
\label{sec:comets}

\subsection{Historical introduction}

Objects classified as comets display the widest variety of dynamical
behaviour in the Solar system. Unfortunately, there is little or no
agreement as to the precise definition of different cometary types.
Historically, the major cometary division has been between those
classified as `long-period' ($P\!> \! 200$\,yr) and `periodic' or
`short-period' ($P\!<\!200\,$yr) comets (e.g.\ Duncan, Quinn \&
Tremaine 1988, Nakamura \& Kurahashi 1998, Levison 1996, Crovisier
2001).  The majority of the short-period group belong to the so-called
Jupiter-family, most of which have very short
periods indeed (a mean value close to 8~years).  For this reason, the
periodic comets are usually divided simply on the basis of orbital
period into two classes: Jupiter-family comets with $P\!<\!20\,$yr
(e.g.\ Fern\'andez 1985, 1994), and Halley-type comets with $20\!<\! P
\! < \! 200\,$yr (e.g.\ Carusi et al.\ 1987b).

Since 1999, however, one-apparition comets with periods in the range
$30\!  < \! P \! < \! 200\,$yr have been given a standard cometary
`C/' designation, rather than the usual periodic `P/' identifier
(Marsden 1999b, Green 2000).  The official range for short-period
comets therefore now extends to only 30\,yr instead of the former
200-year limit, and comets with $30\! < \! P \! < \! 200\,$yr are now
described as `intermediate period' (Marsden \& Williams 1999). The
principal disadvantages of this change are that one-apparition
intermediate-period comets cannot be easily identified
as periodic simply from their designation,
while the phrase `intermediate period' in fact has a rather wide
range of previous connotations, having been used to describe objects
with periods both in the Halley-type $20\! < \! P \!  < \! 200\,$yr
range (e.g.\ Fern\'andez 1980, Fern\'andez \& Gallardo 1994) and with
much longer periods up to $\sim\! 10^3\,$yr (e.g.\ Everhart 1974).

In fact, the situation is more confused still. Whipple (1978) appears
to have chosen $P\! = \! 25\,$yr to separate his `Class~V'
short-period comets (essentially the classical Jupiter-family) from
his Class~IV intermediate-period group ($25\!<\! P \! < \! 1000\,$yr);
while Everhart (e.g.\ 1972, 1973) adopted 13~years instead of 20~years
as a convenient boundary between short-period and intermediate-period
comets, the latter being defined to have periods in the range $13\!<\!
P \! < \!  1000\,$yr. Others (e.g.\ Rickman \& Froeschl\'e 1988, Stagg
\& Bailey 1989, Wetherill 1991) have instead adopted various aphelion
distances $Q$ in the range 8--10\,au to distinguish Jupiter-family
comets from Halley types, while a few authors (e.g.\ Nakamura \&
Kurahashi 1998) have simply considered all comets with $P\! < \!
1000\,$yr as `periodic'.

Contrary to the view expressed by Fern\'andez (1994), therefore, a
classification based on orbital period alone is {\sl not\/} very
useful. The range of definitions used by different investigators to
describe essentially the same cometary sub-groups appears greatly, and
unnecessarily, to complicate any comparison of observational or
theoretical results obtained by different authors. We hope that our
historical summary has convinced the reader that there is a clear need
to improve the taxonomy for comet-like bodies in the Solar system!

\begin{table}
\begin{center}
\begin{tabular}{ccc}\hline
 & & \\
Object & Perihelion & Aphelion \\ 
 & & \\
\hline
 & & \\
E  & $q \lta 4$ & $ Q \lta 4 $ \\ 
SP & $q \lta4$ & $ 4 \lta Q \lta 35 $ \\ 
I  & $q \lta4$ & $ 35 \lta Q \lta 1000$ \\ 
L  & $q \lta4$ & $ Q \gta 1000$ \\
 & & \\
 \hline
 & & \\
J  & $4\lta q\lta6.6$ & $ Q\lta 6.6$ \\ 
JS & $4\lta q\lta6.6$ & $ 6.6\lta Q \lta 12.0$ \\ 
JU & $4\lta q\lta6.6$ & $ 12.0 \lta Q \lta 22.5$ \\ 
JN & $4\lta q\lta6.6$ & $ 22.5 \lta Q \lta 33.5$ \\ 
JE & $4\lta q\lta6.6$ & $ 33.5 \lta Q \lta 60.0$ \\ 
JT & $4\lta q\lta6.6$ & $ Q \gta 60.0$ \\ 
 & & \\
\hline
\end{tabular}
\caption[]
{The upper panel shows the classification scheme for comets.  We
define Encke-type (E), short-period (SP), intermediate-period (I), and
long-period (L) comets based on perihelion and aphelion positions.  The
lower panel shows the scheme for objects whose perihelion is under
Jupiter's control.  The J class describes objects for which both
perihelion and aphelion are under Jupiter's control, the JS class for
which the perihelion is under Jupiter's and the aphelion under Saturn's
control, and so on. In the final two classes, E stands for
Edgeworth-Kuiper belt and T for the trans-Neptunian region immediately
beyond the Edgeworth-Kuiper belt, i.e.\ the `trans-EK belt'.}
\label{table:comets}
\end{center}
\end{table}

\subsection{Taxonomy of comets}

Jupiter's zone of control extends between approximately 4 and 6.6 au. In
our classification scheme, a `comet' has a perihelion $q < 4$ au, and so
lies formally outside Jupiter's zone of control. Nonetheless, Jupiter
dominates the lives of all the short-period comets. For example,
1P/Halley has perihelion within the orbit of Venus and aphelion just beyond
the orbit of Neptune, yet Jupiter plays the most important role in its
evolution (Carusi et al. 1988).  We propose a four-fold division of comets
with perihelion distances less than approximately 4 au, to include Encke (E),
short-period (SP), intermediate-period (I) and long-period (L) types.
The division is
based on aphelion distance, with -- as usual -- further sub-division
into the Tisserand parameter classes (Table~\ref{T1}) denoted by
Roman-numeral subscripts. Further embellishments on this basic scheme
could be introduced to accommodate the resonant and non-resonant main-belt
asteroids (e.g.\ the Hildas, in the 3:2 mean-motion resonance with
Jupiter), and objects on Earth-crossing or Earth-approaching orbits with
periods much less than that of Jupiter, such as near-Earth objects. The
approximate boundaries between the E, SP, I and L comet types are listed
in Table~\ref{table:comets}.

Working outwards from the Sun, we first come to the E-type comets. These have
aphelion closer to the Sun than approximately 4 au, i.e.\ $Q$ within the main
belt.  The prototype is 2P/Encke ($Q \sim 4.1$\,au)
which lies on an orbit that does not
approach closer than a few Hill radii to Jupiter, making it stable within
the inner Solar system for periods of time far greater than that of
typical comets. (The orbit is not, however, stable for indefinitely long
periods, as it is subject to long-term secular perturbations which drive
the perihelion into the Sun on a time-scale of $\sim$100\,000 yr; see
Levison \& Duncan 1994, Farinella et al.\ 1994, Valsecchi et al.\ 1995 for
details.)  Comet 107P/Wilson-Harrington, with a similar orbit, is a 
largely inactive body also known as asteroid (4015), and noted by Marsden
(1992) to be identical to the low-activity comet C/1949\,W1 (=1949g)
following Bowell's identification of precovery observations of the
asteroid on Palomar Schmidt images. Another object, (2201)\,Oljato, has
displayed outgassing in the past, and although currently classified as an
Apollo-type asteroid could well be an evolved Encke-type comet (e.g.\
Weissman et al.\ 1989, McFadden et al.\ 1993).  Many similar bodies in this
class appear to be bona fide asteroids, rocky bodies perhaps originating
via collisions in the main asteroid belt.

Next we come to the SP comets, which include most objects conventionally
thought of as Halley-type and Jupiter-family comets. Many authors have
used the Tisserand parameter with respect to Jupiter to separate these
classes. Levison (1996) suggested that comets with $T_J\! < \! 2.0$ should
be defined as Halley-types and those with $T_J\!  > \! 2.0$ should be
Jupiter-family. However, the precise value of $T_J$ separating the two
types of behaviour is somewhat arbitrary and may depend on period.  For
example, Rickman \& Froeschl\'e (1988) and Bailey (1992) identified a
value of the order of 2.5 as a better marker of a dynamically significant
separation.  Rather than worry about the precise boundary between
Jupiter-family and Halley-types, we use the four-fold division according
to Tisserand parameter introduced in Section~\ref{sec:tisserand}, namely,
Class I with $\TJ \lta 2.0$, Class II with $2.0 \lta \TJ \lta 2.5$, Class
III having $2.5 \lta \TJ < 2.8$ and Class IV having $\TJ \gta 2.8$. Very
roughly, these can be thought of as corresponding to Halley-types,
transition-types, loosely bound Jupiter-family comets and tightly bound
Jupiter-family comets respectively.  We call all objects with $q < 4$ au
and $4< Q <35$ au short-period (SP) objects, and differentiate between
objects within the SP class using the Tisserand parameter with the
sub-class denoted by a Roman numeral subscript.  So, for example,
an SP$_{\rm I}$ object corresponds to a Halley-type comet under the
conventional classification.

The third class consists of I-type comets, which have perihelion distances
less than 4 au and aphelia between approximately 35 and 1000 au. Comets
with aphelion distances less than about 35 au frequently librate around
high-order resonances with Jupiter (Carusi et al.\ 1987a,b, 1988; Chambers
1997) and show evidence for systematic secular changes of their orbital
elements (Bailey \& Emel'yanenko 1996).  Recent examples of I-type comets
are C/1995 O1 (Hale-Bopp) and 153P/Ikeya-Zhang. They have Tisserand
parameters with respect to Jupiter of 0.879 and 0.040 respectively, and so
both fall into the the I$_{\rm I}$ category.

The fourth class is the L-type comets, which have perihelion distances
less than 4 au and aphelia greater than 1000 au. Comets classified under
the conventional system as long-period fall into either the L or the I
class in our system.  L-types -- such as C/1910\,A1 (the Great January
comet) -- have made no more than their first few passages through the
inner Solar system. By contrast, I-types are expected to have made rather
more perihelion passages on average. It is sometimes useful to sub-divide
L-types into new and young (e.g.\ Fern\'andez 1981; Wiegert \& Tremaine
1999; Horner \& Evans 2002). New comets, with $a  \gta 10\,000$ au, are
first time entrants to the Solar system from  the Oort Cloud.

We are of course aware that many comets display activity beyond 4 au from
the Sun, and similarly that there is an increasing number of known objects
with perihelion distances beyond 4 au that have been classified as
active comets (e.g.\ C/2002\,V2 [LINEAR] and C/2003\,A2
[Gleason]) or which have distinctly cometary orbital characteristics (e.g.\
2002\,VQ$_{94}$).  We seek to develop a coherent classification scheme
that is useful for all comet-like bodies in the Solar system.

From now on, as we move outwards in perihelion, we are able to exploit the
idea that comet-like bodies lie under the control of the giant planets.
After the `comets', the next class of objects we encounter are those whose
perihelion lies within the zone of control of Jupiter.  In the
conventional scheme, these bodies are still classified either as
Jupiter-family comets, or as asteroids if they show no outgassing. In our
picture, the designations are listed in Table~\ref{table:comets}. Here,
and henceforth, our classification scheme proceeds with the first letter
designating the planet controlling the perihelion and the second letter
designating the planet controlling the aphelion or the region in which the
aphelion lies. The Tisserand parameter is defined with respect to the
planet which controls the perihelion. So, for instance, a JN$_{\rm IV}$
object has a perihelion under Jupiter's control, an aphelion between 22.5
and 33.5 au and so under Neptune's control, and a Tisserand parameter with
respect to Jupiter $\gta 2.8$; while a JE$_{\rm III}$ type has an aphelion
between 33.5 and 60 au and so lying in the Edgeworth-Kuiper belt and a
Tisserand parameter with respect to Jupiter in the range 2.5 to 2.8.

An example of a J$_{\rm IV}$ object is 29P/Schwassmann-Wachmann\,1,
which has a near-circular orbit just beyond Jupiter.  Its Tisserand
parameter is $\TJ = 3.0$ and so it is not quite Jupiter-crossing. The
object itself is unusual in a couple of respects. The low eccentricity
of its orbit coupled with its unusually great size ($\sim$20\,km
radius) allows it to be followed with ease through its entire
orbit. It undergoes regular outbursts, with its visual magnitude
rising from its usual 17 (at perihelion) by a number of magnitudes,
reaching as high as 10 at its brightest. 

An example of a JS$_{\rm IV}$ type object is 39P/Oterma. It was originally
discovered in 1942 orbiting in a Hilda-type 3:2 mean-motion resonance with
Jupiter, with a low-eccentricity orbit and a period of a mere 8 years. At
that time, its orbit lay between those of Mars and Jupiter and it would
have been classified as an SP$_{\rm IV}$ comet in our picture (although
clearly as a resonant `Hilda' in a more complete scheme). However, it was
later noted that it had entered the 3:2 resonance in 1937 from the 2:3
resonance, and returned to the 2:3 in 1963 (e.g., Kazimirchak-Polanskaya
1972).  Continuous periods in any one resonance are short owing to the
orbit's unstable nature, but it is noteworthy that $\TJ \simeq 3.0$ was
approximately conserved during this evolution. The present orbit lies
outside that of Jupiter, with a current perihelion distance of 5.46 au and
an aphelion of 9.02 au.  This object seems to be a good example of a
recent `handing down' of a Centaur from Saturn's control to Jupiter, with
further periods as an SP comet seeming quite probable.  Finally, recent
discoveries like C/2002 A1 (LINEAR) and C/2002 A2 (LINEAR), with
semi-major axes close to that of Uranus, furnish examples of JN$_{\rm
III}$ objects.

%

\subsection{Resonances}

A key feature of the orbital evolution of Centaurs and many other
comet-like objects in the outer Solar system is the tendency for the
semi-major axis to lie close to one or another mean-motion resonance with
a planet.  This behaviour is exemplified by the so-called Plutinos
(objects in the 2:3 mean-motion resonance with Neptune, i.e.\ around $a \!
\sim \! 39.5$ au), which appear clearly in Fig.~\ref{fig:centaurpop},
and also by the evolution of 2000\,FZ$_{53}$ shown
in the top left plot of Fig.~\ref{fig:centaurone}.  The occurrence
or otherwise of a long-lived resonance will clearly affect the transfer
probability from one part of the $(q,Q,\TP)$ phase space to another and is
an important consideration in the dynamical evolution of outer
Solar system objects. There are recently discovered objects with
semi-major axes close to essentially all the significant Neptunian
resonances, from 1:1 [e.g.\ 2001\,QR$_{322}$ and 2002\,PN$_{34}$] to 1:8
[e.g.\ (54520) 2000\,PJ$_{30}$].  However, detailed discussion of the
classification and role of such resonances is beyond the scope of this paper.

%
%
\begin{table}
\begin{center}
\begin{tabular}{ccc}
\hline
 & & \\
Object	& Perihelion & Aphelion \\ \hline
 & & \\
S      & $6.6 \lta q\lta 12.0$     & $Q \lta  12.0$  \\ 
SU     & $6.6\lta  q \lta 12.0$    & $12.0 \lta  Q \lta  22.5$  \\ 
SN     & $6.6\lta  q \lta 12.0$    & $22.5 \lta  Q \lta  33.5$  \\ 
SE     & $6.6\lta  q \lta 12.0$    & $33.5 \lta  Q \lta  60.0$  \\ 
ST     & $6.6\lta  q \lta 12.0$    & $Q \gta 60.0 $  \\ 
U      & $12.0\lta  q \lta 22.5$   & $Q \lta  22.5$  \\ 
UN     & $12.0\lta  q \lta  22.5$  & $22.5  \lta  Q \lta  33.5$  \\ 
UE     & $12.0\lta  q \lta  22.5$  & $33.5 \lta  Q \lta  60.0$  \\ 
UT     & $12.0\lta  q \lta  22.5$  & $Q \gta  60.0$  \\ 
N      & $22.5\lta  q \lta  33.5$  & $Q \lta  33.5$ \\ 
NE     & $22.5\lta  q \lta  33.5$  & $33.5 \lta  Q \lta  60.0$ \\ 
NT     & $22.5\lta  q \lta  33.5$  & $Q \gta  60.0$  \\ 
EK     & $33.5\lta  q \lta  60.0$  & $Q \lta  60.0$  \\ 
T      & $33.5\lta  q \lta  60.0$  & $Q \gta  60.0$  \\ 
 & & \\
\hline
\end{tabular}
\caption[] {Object classification for the trans-Jovian region. The first
letter designates the planet controlling the perihelion, the second letter
the planet controlling the aphelion or the region in which the aphelion
lies, with the final two classes EK and T being beyond all the giant
planets. (S = Saturn, U = Uranus, N = Neptune, EK = Edgeworth-Kuiper belt,
T = trans-EK belt).}
\label{table:centaurs}
\end{center}
\end{table}

\begin{figure*}
\vspace{2cm}
Available as a gif file (Currentobs1.gif)
\vspace{2cm}
\caption[]{Plot of eccentricity versus semi-major axis showing the
boundaries of the classification scheme. The red circles are listed as
Centaurs and Scattered Disk Objects by the Minor Planet Center, while the
green circles are Edgeworth-Kuiper belt members.  In both cases, we only
record objects if their observed arcs are 30 days or more.}  
\label{fig:centaurpop}
\end{figure*}
\begin{figure*}
\vspace{2cm}
Available as a gif file (FZ53.gif)
\vspace{2cm}
\caption{The evolution of dynamical quantities for a clone of the Centaur
2000 FZ$_{53}$.  The left panel of plots show the variation of semi-major
axis, perihelion, aphelion, eccentricity, inclination and Tisserand
parameter (in this case with respect to Uranus) against time in years. 
The rightmost plot shows the evolution in the plane of semi-major axis and
eccentricity, with the different categories of the classification scheme
marked. The clone spends most of the first 1\,Myr as a UE object and ends
in the NE category. Notice the very sharp transition from UN to UE object
at $\sim 2$ Myr. [Starting orbital elements of the clone are $a=23.595$
au, $e=0.497$, $i= 34.865^\circ$ at JD 2451120.5.]}
\label{fig:centaurone}
\end{figure*}
\begin{figure*} 
\vspace{2cm}
Available as a gif file (Chiron.gif)
\vspace{2cm}
\caption[]{The evolution of dynamical quantities for a clone of 
Chiron. The Tisserand parameter plotted here is that with respect to
Jupiter. The clone starts as an SU object, evolves into a short-period
Jupiter-family comet and then is finally ejected from the Solar
system. Notice how evolution often proceeds along lines of almost
constant aphelion or perihelion in the rightmost panel. [Starting
orbital elements of the clone are $a = 13.599$ au, $e =0.384$,
$i=6.879^\circ$ at JD 2451220.5.]}
\label{fig:centaurtwo}
\end{figure*} 
\begin{figure*}
\vspace{2cm}
Available as a gif file (TF35.gif)
\vspace{2cm} 
\caption[]{The evolution of dynamical quantities for a clone of 1998
TF$_{35}$. The Tisserand parameter plotted here is that with respect
to Jupiter.  The clone starts as a UE object, evolves into a
short-period Jupiter-family comet and then is ejected from the Solar
system. [Starting orbital elements of the clone are $a=26.412$,
$e=0.369$, $i=12.564^\circ$ at JD 2451220.5.]}
\label{fig:centaurthree} 
\end{figure*}
\begin{figure*}
\vspace{2cm}
Available as a gif file (Comets.gif)
\vspace{2cm}  
\caption[]
{The evolution of comets P/1997 T3 (Lagerkvist-Carsenty), upper panels,
and P/1998 U3 (J\"ager), lower panels. On the left, the numerically
integrated orbit is shown together with the orbits of Jupiter, Saturn
and Uranus. On the right, the evolution of the comet in the plane of
semi-major axis and eccentricity is presented. We have suppressed some
of the labels to reduce notational clutter. Notice that
P/1997 T3 shows a perihelion-aphelion exchange on
transference from the Saturn dominated r\'egime to Jupiter family. For
comet J\"ager, the evolution proceeds through the encounter at almost
constant aphelion.}
\label{fig:lagerkvist}
\end{figure*}

\section{Centaurs and trans-Neptunian objects} 
\label{sec:cent}

Table~\ref{table:centaurs} gives the classification scheme for objects
under the control of giant planets other than Jupiter. As before, the
notation is that the first letter designates the planet controlling the
perihelion, and the second letter the planet controlling the aphelion or
the region in which the aphelion lies.  So, a UN object has a perihelion
controlled by Uranus and an aphelion controlled by Neptune, and so on. The
letters EK denote objects with perihelia and aphelia lying close to or
within the classical Kuiper or Edgeworth-Kuiper belt.  Trans-EK belt or T
objects, on the other hand, may have much larger aphelion distances, and
(perhaps depending on their perihelion distances) may be stable `outer'
objects as described by Emel'yanenko et al.\ (2003), or dynamically active
objects similar to those conventionally known as scattered disc objects. 
Both EK and T classes have perihelia beyond Neptune's control (33.5 au)
but less than the conventional outer edge of the Edgeworth-Kuiper belt
($\sim$ 60 au).  They differ in that the aphelion $Q$ is less than
approximately 60 au for the former and greater than 60 au for the latter.
As for the comets, the information on the zone of control must be
supplemented with the Tisserand parameter with respect to the
controlling planet to give a complete classification. The Tisserand
parameter class is given as a subscript.  It is interesting to note the
recent discovery of several high-inclination Centaurs, for example
2002\,VQ$_{94}$ and 2002\,XU$_{93}$, which we classify as ST$_{\rm I}$ and
UT$_{\rm I}$ respectively.

The first Centaur discovered, (2060) Chiron, has a perihelion distance of
8.4 au and an aphelion of 18.8 au. It is classified as an SU$_{\rm IV}$
object, as its Tisserand parameter $\TS = 2.9$. This is one of the largest
Centaurs and is active at perihelion. Hahn \& Bailey (1990) showed that
Chiron's orbit is rapidly evolving and that it may have been a
short-period comet in the past, and will probably become one again in the
future. The second Centaur discovered, (5145) Pholus, has a perihelion of
8.7 au and an aphelion of 32.1 au, together with a Tisserand parameter
$\TS$ of 2.6. It is classified as an SN$_{\rm III}$ object. Pholus's orbit
is also unstable with a characteristic lifetime of $\sim \! 10^6$ yr.  As
another example, (10199) Chariklo has a perihelion of 13.1 au which is
just within Uranus's control and an aphelion of 18.6 au. Its Tisserand
parameter with respect to Uranus is 2.9, making its classification U$_{\rm
IV}$.  Chariklo's orbit is more stable than either Chiron's or Pholus's,
as the dynamical lifetime is roughly a factor of 10 times longer.  Not all
orbits in this part of the Solar system are evolving. For example, Holman
(1997) found that a few low-eccentricity objects between 24 and 27 au
($\sim$0.3\% of an initial low-eccentricity primordial sample) may survive
for 4.5 Gyr without significant dynamical evolution. These objects lie in
the N$_{\rm IV}$ class.  Others (e.g.\ 2001\,QR$_{322}$, also classified
as N$_{\rm IV}$) are recognized as long-lived Neptune-Trojans (Marsden
2003), and it is probably only a matter of time before similar examples of
Trojans are confirmed around Uranus as well.

Almost all the known Centaurs lie within the Tisserand parameter classes
III and IV. There are only a few exceptions -- for example, 2000 FZ$_{53}$
has a Tisserand parameter with respect to its controlling planet (Uranus)
of 2.4. However, the nearly isotropic influx from the Oort Cloud must
inevitably produce at least a few high-inclination Centaurs. Hence, the
other classes are needed, and their apparent emptiness at present is
largely a selection effect.

Fig.~\ref{fig:centaurpop} shows the plane of semi-major axis and
eccentricity.  The red circles show the current location of the 52 objects
listed as Centaurs and Scattered Disk Objects in 2002 June by the Minor
Planet
Center\footnote{http://cfa-www.harvard.edu/iau/lists/Centaurs.html}. A
further 16 objects lie at semi-major axes too great to be seen on the
plot. These objects were chosen from the full list of Centaurs by
requiring the observable arc to be greater than 30 days.  Objects with
shorter arcs are excluded as their orbits often change substantially from
the original estimates when the arc is extended.  The known population
depends both on the true numbers in the various regions and on discovery
selection effects.  Orbits under the control of Uranus and Neptune are
expected to be more long-lived than those under the control of Saturn and
Jupiter. However, this is counterbalanced by the fact that objects nearer
the Sun are more easily discovered.  The question can be studied further
by modelling the efficiency of observational surveys (Jedicke \& Herron
1997).  For the present Centaur population, there is a suggestion in
Fig.~\ref{fig:centaurpop} of a substantial population in the UE category. 
These may be being scattered down from planet to planet having originated
in the Edgeworth-Kuiper belt.

The green circles in Fig.~\ref{fig:centaurpop} show the current
classification of those 373 objects with arcs greater than 30 days
belonging to the Edgeworth-Kuiper belt.  Again, our data were downloaded
in June 2002 from the Minor Planet
Center\footnote{http://cfa-www.harvard.edu/iau/lists/TNOs.html}. The two
most populous categories are NE and EK. The Plutinos, a population of
objects trapped in the 2:3 mean-motion resonance with Neptune (e.g.\
Jewitt \& Luu 1996), fall into either the NE or EK category depending on
eccentricity.  The resonance acts to prevent close approaches to Neptune,
although it is unlikely that such a mechanism would generally
protect all the objects from eventual encounters with Neptune over
time-scales comparable to the age of the Solar system. The EK category
contains most of what are conventionally called the `classical' Kuiper or
Edgeworth-Kuiper belt population, while the T category contains most of
what are conventionally called scattered disc objects.  These last two
classes can be further subdivided (Emel'yanenko et al.\ 2003).

\section{Numerical Examples}

We now illustrate the usefulness of the classification scheme with
some examples drawn from a suite of numerical simulations. Clones of
the current Centaur population are integrated in the forward and
backward direction for up to 6 Myr using the hybrid integrator in the
{\sc Mercury} program (Chambers 1999). The orbits are integrated under
the gravitational effects of the Sun and the four major planets
only. The time-step is 120 days. Once objects passed beyond 1000\,au,
their orbits are no longer followed.

\subsection{From UE$_{\rm II}$ to NE$_{\rm II}$ object}


Fig.~\ref{fig:centaurone} shows the evolution of a clone of
2000\,FZ$_{53}$.  For most of the first 1 Myr its
perihelion is under the control of Uranus
and its aphelion lies in the Edgeworth-Kuiper belt.  The Tisserand
parameter with respect to Uranus is $\TU =2.4$. The clone starts off as an
SE$_{\rm II}$ object and moves rapidly into the UE$_{\rm II}$ zone.

During the course of the simulation, the clone is driven gradually
towards a stable orbit with perihelion close to Neptune (an NE
object). This illustrates one route by which an object can approach
the Edgeworth-Kuiper belt from the Centaur region.  There are two
sharp transitions visible in Fig.~\ref{fig:centaurone}.  The first
change, at 2.0\,Myr from UN to UE, is caused by a close encounter at
aphelion with Neptune; the second, at 2.9\,Myr within the UE category,
is caused by an encounter at perihelion with Uranus.  At the end of
the 6\,Myr simulation, the object has a perihelion just beyond the
orbit of Neptune, a high inclination, and a low eccentricity. The
value of its Tisserand parameter with respect to Neptune is $\TN =
2.4$.  The orbit is similar to those of a number of objects
currently residing in the inner regions of the Edgeworth-Kuiper belt,
such as 1999\,CP$_{133}$.

The stability of the semi-major axis towards the end of the
2000\,FZ$_{53}$ simulation suggests that it lies in a mean-motion
resonance with Neptune. This is borne out by calculation of the orbital
period of the clone, which is $\sim$202\,yr, suggesting that the object
probably lies in the 4:5 mean-motion resonance with Neptune (period
$\sim$164\,yr).  It is possible that the orbit will continue to drift
outward and stabilize after the end of the simulation, illustrating a
rather rare example of outward rather than inward evolution.  The ultimate
destiny of such an object may therefore be to join the Edgeworth-Kuiper
belt or the region beyond it.

\subsection{From SU$_{\rm IV}$ object to SP$_{\rm IV}$ comet}


Fig.~\ref{fig:centaurtwo} shows the evolution of a clone of Chiron (cf.\
Hahn \& Bailey 1990). This object illustrates one of the supply routes for
the short-period comets. The clone starts with its perihelion under the
control of Saturn, but its aphelion wanders between the control of Saturn,
Uranus and Neptune respectively for the first $2 \! \times \!  10^4$ yr.
After this, the clone falls under Jupiter's control through a succession
of aphelion-perihelion interchanges.  The Tisserand parameter $\TJ$ falls
below 3 and the clone spends the next part of its lifetime ($\sim 7 \!
\times \! 10^4$\,yr) as a short-period comet before leaving the SP
region. This is longer than the typical fading time ($\sim 1.2 \times
10^4$\,yr) for Jupiter-family comets found by Levison \& Duncan (1997). 
It is noteworthy that this 200 km diameter object has a significant
probability of evolving to an orbit with perihelion distance less than
that of the Earth (Hahn \& Bailey 1990).  This emphasizes the potential
significance for the inner solar system of so-called `giant' comets and
their disintegration products when time-scales of the order of $10^5$\,yr
or more are considered (e.g., Napier 2001). From the characteristic shape
of the trajectory in the rightmost panel of Fig.~\ref{fig:centaurtwo}, we
see that the evolution proceeds along lines of roughly constant perihelion
or aphelion both before and after the object comes under Jupiter's
control.  The part of the clone's life when it is controlled by Jupiter is
characterized by it being highly chaotic and showing rapid dynamical
evolution.

Note that the clone starts with a Tisserand parameter with respect to
Saturn $\TS = 2.9$. During its lifetime as a short-period comet, its
Tisserand parameter with respect to Jupiter $\TJ = 2.8$. Just as in our
previous example, the Tisserand parameter with respect to the controlling
planet is nearly preserved during the whole evolution.

The fact that the clone stays in an area in which it will be active as
a comet for such a long time suggests that cometary bodies can be
captured into the inner Solar system for long enough to have a
reasonable chance of decoupling from Jupiter, whether by
non-gravitational forces such as outgassing (cf.\ Kres\'ak 1982,
Harris \& Bailey 1998, Asher, Bailey \& Steel 2001) or collisional
mechanisms.  Integrations occasionally provide examples of Encke-type
orbits evolving to become Jupiter-family short-period comets, but the
reverse evolution by gravitational means is so rare that it has never
been seen (Carusi \& Valsecchi 1987).  It is interesting to note that
the stability of an orbit approaching Saturn is lower than one
approaching merely Uranus or Neptune. Similarly, an orbit in the
vicinity of Jupiter is yet more unstable. This adds weight to our
argument that classifying all objects in the trans-Jovian region
merely as `Centaurs' is insufficient to explain the full richness of
dynamical behaviour in the region.

\subsection{From UE$_{\rm IV}$ object to SP$_{\rm IV}$ comet}


Fig.~\ref{fig:centaurthree} shows the evolution of a clone of 1998
TF$_{35}$. It starts off as a UE object with Tisserand parameter $\TU =
2.9$.  This object is of interest since it passes through a number of
different categories (UE, UN, NE, N, U, SU and S) before finally becoming
a short-period comet and then suffering ejection from the Solar system. 
Generally, it moves through the categories sequentially rather than
frequently jumping back and forth between classes.   The clone begins in
an orbit influenced both by Neptune and Uranus, and distant encounters
provide gradual perturbations for most of its lifetime. Eventually,
encounters with Neptune move the object into a near-circular orbit
controlled only by that planet. This period of evolution
ends with a close encounter
which moves the clone's perihelion back to the control of Uranus. The
perihelion drops further until the object encounters Saturn, at which
point it is rapidly handed down to Jupiter's control, where it becomes a
short-period comet for a brief time ($\sim \! 10^4$\,yr) before being
ejected from the Solar system. Whilst a short-period comet, $\TJ = 2.9$,
so there is again excellent conservation of the Tisserand parameter with
respect to the controlling planet.

\subsection{Comets Lagerkvist-Carsenty and J\"ager}

As a final example, let us consider two objects conventionally classified
as short-period Jupiter-family comets, namely P/1997\,T3
(Lagerkvist-Carsenty) and P/1998\,U3 (J\"ager). Numerical integrations by
Lagerkvist et al.\ (2000) showed that both objects suffered recent close
encounters with Saturn that drastically changed their orbits.
Fig.~\ref{fig:lagerkvist} shows the evolution of the orbits of both comets
over the last century. Comet Lagerkvist-Carsenty started out as an
SU$_{\rm IV}$ object in January 1900.  A close encounter with Saturn in
October 1954 transferred its osculating elements through almost
instantaneous SN and SE phases into its present JS$_{\rm IV}$ orbit.  This
proceeded through a perihelion-aphelion interchange, with the old
perihelion becoming the new aphelion.  By contrast, Comet J\"ager started
as an S$_{\rm III}$ object in January 1900 and underwent a direct
transition to its present SP$_{\rm II}$ orbit in July 1991.  The evolution
proceeded at nearly constant aphelion, as can be deduced both from the
plot of the orbit and from its evolutionary track in the plane of
semi-major axis and eccentricity. In this case, the close approach caused
the Tisserand parameter class to change, the controlling planet switching
from Saturn to Jupiter. 

These orbital integrations provide illustrations of why the classification
scheme is useful. The scheme is based around the aphelion and perihelion
distances. Close encounters often preserve aphelion or perihelion to
lowest order, or a perihelion-aphelion interchange may take place. Very
often, the Tisserand parameter with respect to the controlling planet is
conserved to a good approximation as well.

\section{Conclusions}

The main aim of the paper is to introduce a new classification system
for comet-like bodies. Minor bodies between Saturn and Neptune are
often described simply as `Centaurs' and those beyond Neptune simply
as `Kuiper Belt Objects'. This is not very enlightening as the
histories and fates of such bodies may be very different. For example,
simulations show that some Centaurs are in long-lived and stable
orbits, whereas others are dynamically highly active and will evolve
rapidly.

The main focus of this paper has been on the Centaurs and short-period
comets.  Our proposition is to classify the comet-like objects beyond
Jupiter according to the planets that control the evolution of their
perihelion and aphelion.  For example, an SN object has a perihelion under
the control of Saturn and an aphelion under the control of Neptune, while
a UE object has a perihelion under the control of Uranus and an aphelion
in the Edgeworth-Kuiper belt.  This provides 20 dynamically distinct
categories of outer Solar system objects in the Jovian and trans-Jovian
regions.  The evolutionary tracks of bodies like Centaurs often show
periods in which the aphelion or perihelion distances are individually
rather well conserved, or encounters in which the old perihelion becomes
the new aphelion or vice versa (`perihelion-aphelion interchanges').

The objects with smaller perihelion distances, which we
designate as `comets', are dominated by
Jupiter and not by the planets nearest at perihelion or aphelion. In our
scheme, comets have perihelion distance less than 4 au. They are
subdivided into Encke-type, short-period, intermediate-period and
long-period according to aphelion distance. Following a succession of
authors beginning with Kres\'ak (1972), we favour
further sub-divisions based on the
Tisserand parameter. Specifically, we subdivide the categories of comets
into: Class I which has $\TJ \lta 2.0$, Class II which has $2.0 \lta \TJ
\lta 2.5$, Class III having $2.5 \lta \TJ < 2.8$ and Class IV having $\TJ
\gta 2.8$. Very roughly, these can be thought of as corresponding to
Halley-types, transition-types, loosely bound Jupiter-family comets and
tightly bound Jupiter-family comets respectively.  This idea is then
extended to all comet-like bodies by identifying the Tisserand parameter
of the planet controlling the perihelion with a subscript.

Given the aphelion and perihelion distance, together with the
Tisserand parameter of the planet controlling the perihelion, it is
straightforward to find the instantaneous classification of any Solar
system object. Our new classification scheme extends the existing
taxonomy for comets to cover all comet-like bodies in the Solar
system.

\section*{Acknowledgments}
This research was supported by the Particle Physics and Astronomy
Research Council (JH), the Royal Society (NWE) and the Northern
Ireland Department of Culture, Arts and Leisure (MEB, DJA). DJA
acknowledges the hospitality of the Japan Spaceguard Association during
work on this paper. We thank the referee for very constructive comments.


\label{lastpage}

\begin{thebibliography}{99}

\item[] Applegate J.H., Douglas M.R., G\"ursel Y., Sussman G.J., Wisdom, J.,
1986, 
AJ, 92, 176

\bibitem[Arnold, Kozlov \& Neishtadt 1987]{arnold} Arnold V.I., Kozlov
V.V., Neishtadt A.I., 1987 in Arnold V.I., ed., Dynamical Systems III.
Springer Verlag, New York, p.~70

\bibitem[Asher, Bailey, Hahn, \& Steel(1994)]{1994MNRAS.267...26A}
Asher D.J., Bailey M.E., Hahn G., Steel D.I., 1994, MNRAS, 267, 26

\bibitem[Asher, Bailey \& Steel (2001)]{abs}
Asher D.J., Bailey M.E., Steel D.I., 2001,
in Marov M., Rickman H., eds, Astrophys.\ Space Sci.\ Libr.\ 261,
Collisional Processes in the Solar System. Kluwer, Dordrecht, p.~121

\item[] Bailey M.E., 1992,
Cel.\ Mech.\ Dyn.\ Astron., 54, 49

\item[] Bailey M.E., Emel'yanenko V.V., 1996,
MNRAS, 278, 1087


\bibitem[Carusi \& Valsecchi 1987]{car} Carusi A., Valsecchi
G.B., 1987, 
European Regional Astronomy Meeting of the IAU, Volume 2,
21

\item[] Carusi A., Kres\'ak \v{L}., Perozzi E., Valsecchi G.B., 1987a,
European Regional Astronomy Meeting of the IAU, Volume 2,
29

\item[] Carusi A., Kres\'ak \v{L}., Perozzi E., Valsecchi G.B., 1987b,
A\&A, 187, 899

\bibitem[Carusi, Valsecchi, Kresak, \& Perozzi 1988]{halley} Carusi
A., Valsecchi G.B., Kresak L., Perozzi E., 1988, Cel. Mech., 43, 319

\bibitem[Chambers(1997)]{chambers97} Chambers J.E., 1997,
Icarus, 125, 32

\bibitem[Chambers(1999)]{1999MNRAS.304..793C} Chambers J.E., 1999,
MNRAS, 304, 793

\bibitem[Charnoz, Th{\' e}bault, \& Brahic 2001]{charn} Charnoz S.,
Th{\' e}bault P., Brahic A., 2001, A\&A, 373, 683

\item[] Crovisier J., 2001,
in Murdin P., ed., Encyclopedia of Astronomy and Astrophysics. IOP
Publishing Ltd and Nature Publishing Group, p.~446

\item[] Dones L., Levison H.F., Duncan M., 1996,
in Rettig T.W., Hahn J.M., eds, ASP Conf.\ Ser.\ 107, Completing the
Inventory of the Solar System.  ASP, San Francisco, p.~233

\bibitem[Duncan \& Levison]{dl} Duncan M.J., Levison H.F., 1997,
Sci, 276, 1670

\item[] Duncan M., Quinn T., Tremaine S., 1988,
ApJ, 328, L69

\bibitem[Emelyanenko et al. 2002]{em} Emel'yanenko V.V., Asher D.J.,
Bailey M.E., 2003, 
MNRAS, 338, 443

\item[] Everhart E., 1972,
Astrophysical Letters, 10, 131

\item[] Everhart E., 1973, 
AJ, 78, 329

\item[] Everhart E., 1974,
in Cristescu C., Klepczynski W.J., Milet B., eds,
Proc.\ IAU Coll.\ 22, Asteroids, comets, meteoric matter, p.~223

\item[] Everhart E., 1977, 
in Delsemme A.H., ed., Proc.\ IAU Coll.\
39, Comets Asteroids
Meteorites: Interrelations, Evolution and Origins. University of Toledo,
Toledo, p.\ 99

\item[] Farinella P., Froeschl\'e Ch., Froeschl\'e C., Gonczi R., Hahn G.,
Morbidelli A., Valsecchi G.B., 1994,
Nat, 371, 314

\item[] Fern\'andez J.A., 1980
MNRAS, 192, 481

\item[] Fernandez J.A., 1981, 
A\&A, 96, 26 

\item[] Fern\'andez J.A., 1985, 
Icarus, 64, 308

\item[] Fern\'andez J.A., 1994,
in Milani A., Di Martino M., Cellino A., eds, Proc.\ IAU Symp.\ 160, 
Asteroids, Comets, Meteors 1993. Kluwer, Dordrecht, p.~223

\item[] Fern\'andez J.A., Gallardo T., 1994,
A\&A, 281, 911


\item[] Fern\'andez J.A., Gallardo T., Brunini A., 2002,
Icarus, 159, 358



\bibitem[Green 2000]{green2000}
Green D.W.E., 2000, International Comet Quarterly, 22, 2

\item[] Hahn G., Bailey M.E., 1990,
Nat, 348, 132

\item[]Harris N.W., Bailey M.E., 1998,
MNRAS, 297, 1227

\item[] Holman M.J., 1997,
Nat, 387, 785

\bibitem[Horner \& Evans(2002)]{he2002}
Horner J., Evans N.W., 2002, MNRAS, 335, 641

\bibitem[Jedicke \& Herron(1997)]{1997Icar..127..494J} Jedicke R.,
Herron J.D., 1997, Icarus, 127, 494

\bibitem[Jewitt \& Luu 1993]{1993Natur.362..730J} Jewitt D., Luu
J.X., 1993, Nat, 362, 730

\bibitem[Jewitt \& Luu(1996)]{1996ciss.conf..255J} Jewitt D., Luu J.X.,
1996, in Rettig T.W., Hahn J.M., eds, ASP Conf.\ Ser.\ 107, Completing
the Inventory of the Solar System.  ASP, San Francisco, p.~255

\item[] Kazimirchak-Polonskaya E.I., 1972,
in Chebotarev G.A., Kazimirchak-Polonskaya E.I., Marsden B.G., eds,
Proc.\ IAU Symp.\ 45, The Motion, Evolution of Orbits, and Origin of
Comets.  Reidel, Dordrecht, p. 373

\bibitem[Kowal 1979]{ctk} Kowal C.T., 1979,
in Gehrels T., ed., Asteroids.  University of Arizona Press, Tucson,
p.~436


\item Kresak \v{L}., 1972,
in Chebotarev G.A., Kazimirchak-Polonskaya E.I., Marsden B.G., eds,
Proc.\ IAU Symp.\ 45,
The Motion, Evolution of Orbits, and Origin of Comets.
Reidel, Dordrecht, p.~503

\item[] Kres\'ak \v{L}., 1980,
The Moon and Planets, 22, 83

\item[] Kres\'ak \v{L}., 1982,
in Fricke W., Teleki G., eds, Sun and Planetary System.  Reidel,
Dordrecht, p.~361

\item[] Kres\'ak \v{L}., 1983,
in West R.M., ed., Highlights of Astronomy, 6, 377

\item[] Kres\'ak \v{L}., 1985,
in Carusi A., Valsecchi G.B., eds, Proc.\ IAU Coll.\ 83, Dynamics of
Comets: Their Origin and Evolution.  Reidel, Dordrecht, p.~279

\bibitem[Larsen et al. 2001]{jalarsen} Larsen J.A. et al., 2001,
AJ, 121, 562

\bibitem[Lagerkvist et al 200]{lhkc} Lagerkvist C.-I., Hahn G., Karlsson
O., Carsenty U., 2000, A\&A, 362, 406

\bibitem[Levison 1996]{levison} Levison H.F., 1996,
in Rettig T.W., Hahn J.M., eds, ASP Conf.\ Ser.\ 107, Completing the
Inventory of the Solar System.  ASP, San Francisco, p.~173

\item[] Levison H.F., Duncan M.J., 1994,
Icarus, 108, 18

\bibitem[Levison \& Duncan 1997]{mh} Levison H.F., Duncan M.J., 1997,
Icarus, 127, 13

\bibitem[Luu \& Jewitt 1990]{luu} Luu J.X., Jewitt D., 1990, AJ, 100,
913

\bibitem[Luu et al.(1997)]{1997Natur.387..573L} Luu J.X., Jewitt D., 
Trujillo C.A., Hergenrother C.W., Chen J., Offutt W.B., 1997,
Nat, 387, 573 

\item[] Manara A., Valsecchi G.B., 1991,
A\&A, 249, 269

\bibitem[Marsden 1992]{marsden} Marsden B.G., 1992, IAU Circ.\
5585

\item[] Marsden B.G., 1999a,
http://cfa-www.harvard.edu/cfa/ps/press\-info/200TNOs.html

\item[] Marsden B.G., 1999b,
Minor Planet Circ.\ 35155

\item[] Marsden B.G., 2003,
Minor Planet Elect.\ Circ.\ 2003-A55

\item [] Marsden B.G., Williams, G.V., 1999, 
Catalogue of Cometary Orbits 1999.
Minor Planet Center, Cambridge, Massachusetts

\bibitem[McFadden et al.(1993)]{1993JGR....98.3031M} McFadden L.A.,
Cochran A.L., Barker E.S., Cruikshank D.P., Hartmann W.K., 1993,
J. Geophys. Res., 98, 3031


\bibitem[Murray \& Dermott]{md} Murray C.D., Dermott S.F., 1999, Solar
System Dynamics. Cambridge University Press, Cambridge

\item[] Nakamura T., Kurahashi, H., 1998,
AJ, 115, 848

\bibitem[Napier 2001]{bill} Napier W.M. 2001, MNRAS, 321, 463

\bibitem[Quinn, Tremaine, \& Duncan(1990)]{1990ApJ...355..667Q} Quinn
T., Tremaine S., Duncan M., 1990, ApJ, 355, 667


\item[] Rickman H., Froeschl\'e Cl., 1988,
Cel.\ Mech., 43, 243

\bibitem[Scotti 1992]{scotti} Scotti J.V., 1992, IAU Circ.,
5434

\item[] Stagg C.R., Bailey M.E., 1989, 
MNRAS, 241, 507

\item[] Stern A., Campins, H., 1996,
Nat, 382, 507

\item[] Vaghi S., 1973a,
A\&A, 24, 107

\item[] Vaghi S., 1973b,
A\&A, 29, 85

\item[] Valsecchi G.B., Morbidelli A., Gonczi R., Farinella P.,
Froeschl\'e Ch., Froeschl\'e C., 1995,
Icarus, 118, 169


\item[] Weissman P.R., 2001,
in Murdin P., ed., Encyclopedia of Astronomy and Astrophysics, IOP
Publishing Ltd and Nature Publishing Group, p.~1930

\item[] Weissman P.R., A'Hearn M.F., McFadden L.A., Rickman H., 1989, 
in Binzel R.P., Gehrels T., Matthews M.S., eds, Asteroids II,
University of Arizona Press, Tucson, p. 880

\item[] Wetherill G.W., 1991,
in Newburn R.L. Jr, Neugebauer M., Rahe J., eds, Proc.\ IAU Coll.\
116, Comets in the Post-Halley Era, Vol.\ 1. Kluwer, Dordrecht,
p.~537

\item[] Whipple F.L., 1978,
The Moon and Planets, 18, 343

\bibitem[Wiegert \& Tremaine(1999)]{1999Icar..137...84W} Wiegert P.,
Tremaine S., 1999, Icarus, 137, 84

\end{thebibliography}
\end{document}